\documentclass[useAMS]{mn2e}
\usepackage{graphicx}

\def\arcsec{\tt ''}
\def\simgt{\ {\raise-.5ex\hbox{$\buildrel>\over\sim$}}\ }

\def\cd{cd$^{-1}$\,}
\def\kms{kms$^{-1}$\,}

\begin{document}

\title{The rapidly oscillating Ap star HD 99563 and its distorted dipole 
pulsation mode}
\author[G. Handler et al.]
{G. Handler,$^{1,2}$ W. W. Weiss,$^{1}$ R. R. Shobbrook,$^{3}$ E.  
Paunzen,$^{1}$ A. Hempel,$^{4}$ \and S. K. Anguma,$^{5}$ P. C.  
Kalebwe,$^{6}$ D. Kilkenny,$^{2}$ P. Martinez,$^{2}$ M. B. Moalusi,$^{7}$ 
\and R. Garrido,$^{8}$ R. Medupe$^{2}$
        \and \\
$^{1}$ Institut f\"ur Astronomie, Universit\"at Wien,
T\"urkenschanzstra\ss e 17, A-1180 Wien, Austria\\
$^{2}$ South African Astronomical Observatory, P.O. Box 9, Observatory
7935, South Africa\\
$^{3}$ Research School of Astronomy and Astrophysics, Australian National
University, Canberra, ACT, Australia\\
$^{4}$ Observatoire de Gen\`{e}ve, Chemin des Maillettes 51, CH-1290
Sauverny, Switzerland\\
$^{5}$ Department of Physics, Mbarara University of Science and 
Technology, P. O. Box 1410, Mbarara, Uganda\\
$^{6}$ Physics Department, The University of Zambia, P. O. Box 32379, 
Lusaka, Zambia\\
$^{7}$ Department of Physics, University of the North-West, Private Bag 
X2046, Mmabatho 2735, South Africa\\
$^{8}$ Instituto de Astrofisica de Andalucia, Apt. 3044, E-18080 Granada,
Spain}

\date{Accepted 2005 nnnn nn.
      Received 2005 nnnn nn;
      in original form 2005 nnnn nn}

\maketitle

\begin{abstract}

We undertook a time-series photometric multi-site campaign for the rapidly
oscillating Ap star HD 99563 and also acquired mean light observations
over two seasons. The pulsations of the star, that show flatter light
maxima than minima, can be described with a frequency quintuplet centred
on 1557.653~$\mu$Hz and some first harmonics of these. The amplitude of
the pulsation is modulated with the rotation period of the star that we
determine with $2.91179 \pm 0.00007$~d from the analysis of the stellar
pulsation spectrum and of the mean light data. We break the distorted
oscillation mode up into its pure spherical harmonic components and find
it is dominated by the $\ell=1$ pulsation, and also has a notable $\ell=3$
contribution, with weak $\ell=0$ and 2 components. The geometrical
configuration of the star allows one to see both pulsation poles for about
the same amount of time; HD 99563 is only the fourth roAp star for which
both pulsation poles are seen and only the third where the distortion of
the pulsation modes was modelled. We point out that HD 99563 is very
similar to the well-studied roAp star HR 3831. Finally, we note that the
visual companion of HD 99563 is located in the $\delta$ Scuti instability
strip and may thus show pulsation. We show that if the companion was
physical, the roAp star would be a 2.03~$M_{\odot}$ object, seen at a
rotational inclination of $44\degr$, which then predicts a magnetic
obliquity $\beta=86.4\degr$.

\end{abstract}


\section{Introduction}

Where the lower part of the classical instability strip intersects the
main sequence, three distinct classes of multiperiodically pulsating
variables can be found. The $\gamma$~Doradus stars pulsate in gravity
modes of high radial order and have periods of the order of one day (Kaye
et al.\ 1999). The $\delta$ Scuti stars have periods of the order of a few
hours (Breger 1979) and are thus pressure and gravity mode pulsators of
low radial order. 

The fastest pulsations in this domain in the HR diagram are however
excited in the rapidly oscillating Ap (roAp) stars (Kurtz 1982, Kurtz \&
Martinez 2000), with typical periods around 10 minutes, indicating
pressure modes of high radial overtones. The photometric semi-amplitudes
associated with these pulsations are in most cases only a few mmag, which
makes them difficult to detect. The roAp stars are also remarkable due to
their spectral peculiarities, since they are pulsating representatives of
cool magnetic Ap stars of the SrCrEu subtype.

The magnetic fields of Ap stars cause elemental segregation on the stellar
surface. In other words, the chemical elements are arranged in patches on
the stellar surface (see, e.g., Kochukhov et al.\ 2004 or Lueftinger et
al.\ 2003 for well-documented examples). This causes modulation of the
mean apparent brightness of the star with the rotation period. As the
chemical elements show some alignment with the magnetic poles, the
rotational light curves show a single-wave structure if only one magnetic
pole is seen and a double-wave variation if both magnetic poles come into
view during a rotation cycle.

The strong magnetic fields present in the roAp stars are usually not
aligned with the stellar rotation axis. Since the pulsation axis does
however coincide with the magnetic axis, the roAp stars are oblique
pulsators (Kurtz 1982). This means that the pulsation modes excited in the
roAp stars are seen at different aspects during the rotation cycle, which,
for nonradial modes, causes amplitude modulation over the rotation period
(which can then be inferred) and allows us to put constraints on the
geometry of the pulsator.

The oblique pulsator model (OPM), in its simplest form, then predicts that
(except for special geometric orientations of the axes) dipole ($\ell=1$)
pulsation modes will be split into equally spaced triplets. Quadrupole
($\ell=2$) modes will give rise to equally spaced frequency quintuplets,
where the spacing of consecutive multiplet components is exactly the
stellar rotational frequency.

The magnetic field of the roAp stars affects the pulsations in two
additional ways. Firstly, the pulsation frequencies are shifted with
respect to their unperturbed value (Cunha \& Gough 2000) and secondly, the
pulsation modes are distorted so that a single spherical harmonic can no
longer describe them fully (e.g., Kurtz, Kanaan \& Martinez 1993, Takata
\& Shibahashi 1995). This is observationally manifested by the presence of
additional multiplet components surrounding the first-order singlets,
triplets or quintuplets, which are spaced by integer multiples of the
rotation frequency.

For a long time, the predominant observing method for studying the
pulsations of roAp stars was time-resolved high-speed photometry, yielding
interesting results on the pulsational behaviour, geometry and
asteroseismology of these stars (e.g., see Matthews, Kurtz \& Martinez
1999). The most recent observational advances however came from
time-resolved spectroscopy: since the vertical wavelengths of the
pulsation modes are comparable to the size of the line-forming regions in
the atmospheres of these stars, the vertical structure of the atmospheres
can be resolved (Ryabchikova et al.\ 2002, Kurtz, Elkin \& Mathys 2003).

Spectroscopy is also more sensitive to the detection of roAp star 
pulsations than photometry. Due to the vertical stratification of chemical 
elements in their atmospheres (e.g.\ Ryabchikova et al.\ 2002) the 
pulsational radial velocity amplitudes of some spectral lines (most 
notably rare earth elements) can reach several kilometres per second. 

The most extreme example of such high radial velocity amplitudes is
HD~99563 (XY~Crt). This star was photometrically discovered to pulsate by
Dorokhova \& Dorokhov (1998) and was confirmed by Handler \& Paunzen
(1999). Elkin, Kurtz \& Mathys (2005) discovered pulsational radial
velocity variations with semi-amplitudes of up 5~\kms for some Eu{\sc II}
and Tm{\sc II} lines in their time-resolved spectroscopy of this star.
Such high amplitudes are capable of yielding information on the structure
of the atmospheres of Ap stars with the best possible signal to noise.

However, one important piece of information that spectroscopy cannot
supply at this point is detailed knowledge of the stellar pulsation
spectrum. The reason is that the largest telescopes are necessary for
obtaining spectra of the required time resolution and signal to noise.  
However, observing time on these telescopes is sparse. Therefore, lengthy
photometric measurements of these stars on small telescopes are still
required to decipher the pulsational spectra fully.

To this end, we included HD 99563 as a secondary target in a multi-site
campaign originally devoted to the roAp star HD 122970 (Handler et al.\
2002) to be observed at times when the latter star was not yet accessible.
However, HD 99563 turned out to be quite interesting, which is why we
continued to observe it after the original campaign was finished. We also
acquired some mean-light observations of HD 99563 in an attempt to
determine its rotation period. In this paper, we report the results of
these measurements.

\section{Observations and reductions}

\subsection{The high-speed photometry}

Our multi-site time-series photometric observations were obtained with
seven different telescopes at four observatories: the 0.75m T6 Automatic
Photometric Telescope at Fairborn Observatory (FAPT) in Arizona, the 0.5m,
0.75m, 1.0m and 1.9m telescopes at the South African Astronomical
Observatory (SAAO), the 0.6m reflector at Siding Spring Observatory (SSO)  
in Australia and the 0.9m telescope at Observatorio de Sierra Nevada (OSN)
in Spain. Whereas the latter observations consisted of Str\"omgren $uvby$
photometry, the other telescopes used the B filter only. The measurements
are summarised in Table 1.

\begin{table}
\caption[]{Journal of the observations of HD 99563; $N_{40}$ is the number 
of 40-second data bins obtained in the respective night and $\sigma$ is
the rms scatter of the residual light curves after prewhitening and
low-frequency filtering; the remaining columns are self-explanatory.}
\begin{center}
\scriptsize
\begin{tabular}{lcccccc}
\hline
Telescope & Date & Start & Length & $N_{40}$ & $\sigma$ \\
 & (UT) & (UT) & (hr) & & (mmag) \\
\hline
FAPT 0.75m & 18/03 & 04:54:24 & 0.64 & 55 & 1.98 \\
FAPT 0.75m & 19/03 & 03:43:40 & 1.75 & 140 & 1.71 \\
FAPT 0.75m & 20/03 & 03:39:41 & 2.15 & 180 & 1.40 \\
FAPT 0.75m & 23/03 & 03:23:03 & 2.54 & 208 & 1.74 \\
FAPT 0.75m & 24/03 & 03:16:41 & 2.53 & 210 & 2.21 \\
FAPT 0.75m & 25/03 & 04:20:58 & 1.43 & 119 & 1.51 \\
FAPT 0.75m & 26/03 & 03:10:24 & 2.52 & 209 & 1.73 \\
FAPT 0.75m & 27/03 & 03:04:34 & 2.52 & 200 & 1.88 \\
SAAO 0.5m & 27/03 & 18:47:49 & 2.68 & 188 & 1.75 \\
SSO 0.6m & 28/03 & 09:49:16 & 2.01 & 166 & 2.04 \\
SAAO 0.5m & 28/03 & 17:59:55 & 3.36 & 252 & 2.14 \\
FAPT 0.75m & 29/03 & 02:58:32 & 2.52 & 174 & 2.14 \\
SSO 0.6m & 29/03 & 09:20:27 & 3.59 & 271 & 2.21 \\
SAAO 0.5m & 29/03 & 17:56:30 & 3.35 & 255 & 1.87 \\
SAAO 0.5m & 30/03 & 20:31:12 & 0.72 & 59 & 1.85 \\
FAPT 0.75m & 31/03 & 02:47:47 & 2.52 & 210 & 2.24 \\
SSO 0.6m & 31/03 & 09:43:39 & 2.82 & 235 & 1.62 \\
SSO 0.6m & 01/04 & 10:41:08 & 2.18 & 177 & 1.48 \\
SAAO 0.5m & 02/04 & 18:03:43 & 2.96 & 192 & 2.19 \\
SAAO 1.0m & 03/04 & 19:34:44 & 1.29 & 104 & 0.59 \\
SSO 0.6m & 04/04 & 09:18:16 & 2.01 & 146 & 1.80 \\
SAAO 1.0m & 04/04 & 19:15:49 & 1.62 & 137 & 0.65 \\
SSO 0.6m & 05/04 & 11:01:34 & 1.37 & 101 & 1.61 \\
FAPT 0.75m & 09/04 & 02:50:41 & 1.61 & 131 & 2.15 \\
SAAO 0.5m & 09/04 & 17:43:48 & 1.08 & 64 & 2.11 \\
SAAO 1.0m & 09/04 & 18:00:40 & 1.14 & 98 & 1.23 \\
OSN 0.9m & 09/04 & 19:54:16 & 1.61 & 135 & 8.2/2.8/3.0/3.7 \\
SAAO 0.5m & 10/04 & 17:50:42 & 2.59 & 191 & 1.47 \\
SAAO 0.5m & 11/04 & 17:27:29 & 3.64 & 264 & 1.62 \\
OSN 0.9m & 11/04 & 20:19:34 & 4.44 & 309 & 7.5/4.4/4.0/4.5 \\
FAPT 0.75m & 12/04 & 02:55:06 & 1.57 & 127 & 1.71 \\
SAAO 0.5m & 12/04 & 17:28:47 & 6.85 & 465 & 1.79 \\
FAPT 0.75m & 13/04 & 02:53:47 & 1.61 & 135 & 1.56 \\
FAPT 0.75m & 14/04 & 03:44:02 & 0.69 & 57 & 2.90 \\
SAAO 0.5m & 14/04 & 17:09:07 & 4.13 & 305 & 1.87 \\
FAPT 0.75m & 15/04 & 02:44:55 & 1.42 & 120 & 1.65 \\
FAPT 0.75m & 16/04 & 02:57:02 & 1.23 & 102 & 1.60 \\
FAPT 0.75m & 18/04 & 02:52:32 & 0.90 & 76 & 1.62 \\
SSO 0.6m & 23/04 & 13:42:01 & 0.72 & 54 & 2.11 \\
SSO 0.6m & 27/04 & 09:01:18 & 5.04 & 398 & 1.71 \\
SAAO 1.9m & 30/04 & 17:14:35 & 0.69 & 60 & 0.95 \\
SAAO 0.75m & 15/05 & 17:11:06 & 1.61 & 137 & 1.57 \\
SSO 0.6m & 16/05 & 11:59:23 & 0.24 & 23 & 2.03 \\
SAAO 0.75m & 16/05 & 20:50:28 & 0.68 & 58 & 1.24 \\
SAAO 0.75m & 17/05 & 18:28:06 & 1.80 & 155 & 1.50 \\
SAAO 0.75m & 18/05 & 16:59:33 & 3.40 & 257 & 1.43 \\
SAAO 0.75m & 21/05 & 19:09:55 & 2.34 & 206 & 1.50 \\
SSO 0.6m & 22/05 & 12:09:16 & 1.10 & 94 & 1.90 \\
SSO 0.6m & 24/05 & 08:38:42 & 3.72 & 298 & 1.41 \\
SSO 0.6m & 30/05 & 08:20:08 & 1.68 & 139 & 1.12 \\
SAAO 0.5m & 01/06 & 18:09:26 & 2.76 & 222 & 1.20 \\
SAAO 0.5m & 02/06 & 18:37:05 & 1.76 & 143 & 1.26 \\
SAAO 0.5m & 04/06 & 17:14:49 & 3.18 & 253 & 1.60 \\
SAAO 0.5m & 06/06 & 19:58:06 & 0.79 & 59 & 2.34 \\
SAAO 0.5m & 07/06 & 16:38:51 & 2.54 & 183 & 1.73 \\
SAAO 0.5m & 08/06 & 16:41:45 & 3.97 & 286 & 2.78 \\
SAAO 0.5m & 09/06 & 16:36:15 & 4.00 & 296 & 1.59 \\
SAAO 0.5m & 10/06 & 18:23:46 & 1.18 & 82 & 1.32 \\
SAAO 0.5m & 11/06 & 17:40:49 & 2.75 & 202 & 1.56 \\
\hline
Total & & $B$ & 125.49 & 9728 & 1.82 \\
\hline
 & & $uvby$ & 6.05 & 444 & 7.7/3.9/3.7/4.3 \\
\hline
\end{tabular}
\normalsize
\end{center}
\end{table}

The time-series photometry was generally acquired as continuous 10-second
integrations. Large apertures ($>$ 35 arcseconds) were employed to
minimise the contributions of seeing and guiding. The observations of the
target star were interrupted at irregular intervals (depending on the
brightness and position of the Moon) to obtain measurements of sky
background.

Reductions were performed by first correcting for coincidence losses (dead
time correction), then for sky background and extinction. Some
low-frequency filtering to remove residual transparency variations or tube
sensitivity drifts with time scales longer than 30 minutes was applied by
fitting low-order polynomials to the data and by subtracting these.  
Finally the measurements were binned into 40-second integrations, the
times of the observation were converted to Heliocentric Julian Date (HJD)
and all the data were joined into a combined single light curve.

\begin{figure*}
\includegraphics[angle=0,width=180mm,viewport=-10 5 560 500]{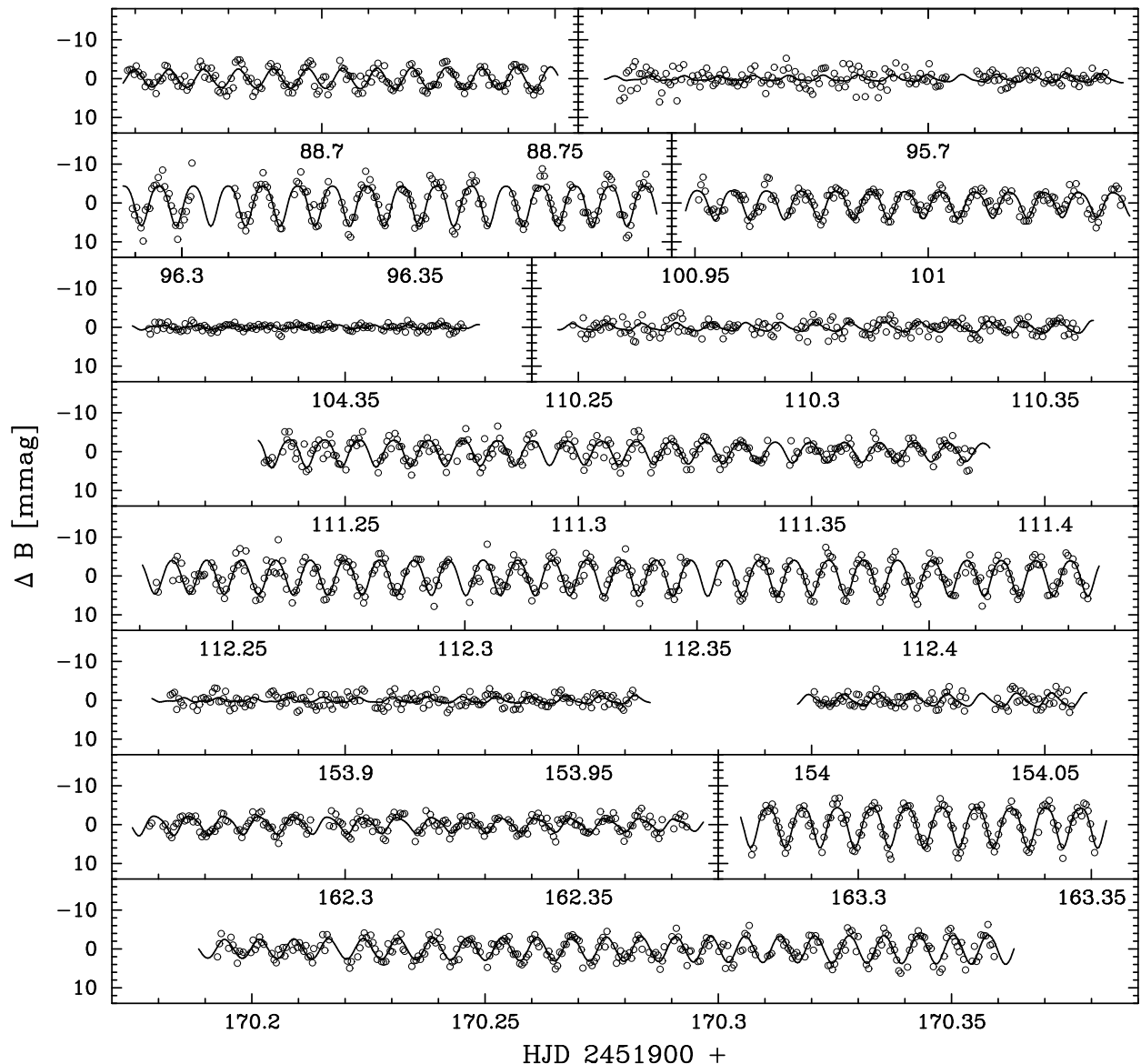}
\caption[]{Some light curves of HD 99563 from our multi-site observations
showing its diverse nightly behaviour. The solid lines represent the 
multifrequency fit to be listed in Table 2, to guide the eye.}
\end{figure*}

Some of our nightly measurements are shown in Fig.\ 1, clearly containing
some interesting features. Firstly, the pulsation amplitude can vary
rather dramatically from almost zero to more than 10 mmag peak-to-peak,
and it can also be seen to vary during a single night. Secondly, the shape
of the light curves near maximum amplitude is peculiar, with the light
maxima being flatter than the light minima.

\subsection{The mean light observations}

Differential multicolour photometry of HD 99563 was obtained with the
0.75m Automatic Photometric Telescope at SAAO (Martinez et al.\ 2002) from
March 2001 until June 2004. Single-channel differential measurements with
respect to the single comparison star HD 99506, which is not known to be
variable, in the Johnson-Cousins UBVRI filters were acquired. Again, large
apertures ($>35\arcsec$) were used to eliminate effects of scintillation
and seeing.

Two nightly UBVRI measurements of HD 99563 were usually taken in between
two measurements of HD 99506, ensuring that variations in sky transparency
can be readily compensated for, but also aiming at averaging out the
effects of the rapid oscillations on the mean magnitude. As the two
integrations on the target were separated by about 380\,s, about half the
oscillation period, the pulsational signal was strongly suppressed.

The data of both stars were reduced by correction for coincidence losses,
sky background and extinction, and the timings were heliocentrically
corrected. Differential magnitudes between the variable and the comparison
star were computed by simple linear interpolation. After rejection of
outliers and merging of the two nightly measurements, a total of 150
differential UBVRI magnitudes of HD 99563 were available.

\section{Frequency analysis (I)}

\subsection{The high-speed photometry}

The reduced B filter light curves were analysed with the computer
programme {\tt Period98} using single-frequency Fourier and
multiple-frequency least-squares techniques (Sperl 1998). This programme
has, amongst others, the capabilities of calculating Fourier periodograms,
calculating the best fit to the data with all given frequencies
simultaneously by minimising the residuals between light curve and fit.  
Within the multifrequency fitting, combination frequencies (that are in
our case expected due to the shape of the light curves) can be fixed to
their expected values and their amplitudes and phases can be optimised
together with the other independent parameters.

We show the amplitude spectrum of the combined light curve in the
uppermost panel of Fig.\ 2. Two main regions of power are visible, one
centered around 1500 $\mu$Hz and a weaker one around 3100 $\mu$Hz. The
first of the sequence of panels on the left-hand side of Fig.\ 2 shows the
spectral window of the data at an extended scale, computed as the Fourier
transform of a single noise free sinusoid with a frequency of
1553.67~$\mu$Hz and an amplitude of 2.7~mmag. We see some alias patterns
about 34.7~$\mu$Hz away from the central peak. This is the 3~\cd alias,
which is most prominent in our data because of the longitude distribution
of the observatories used combined with the short duration of most of the
runs. Fortunately, neither this alias, nor any other, will have any
influence on our results.

The second panel on the left-hand side contains the amplitude spectrum of
the data. It is clearly more complicated than the spectral window. 
Prewhitening the strongest signal from the data we end up with the 
residual amplitude spectrum in the third (middle) panel of Fig.\ 2, 
left-hand side. Further prewhitening of the strongest signal there reveals 
another one, and continuing this procedure results in the detection of a 
fourth frequency in this domain, where no more signals can now be 
detected.

We continue the frequency analysis in the region around 3100~$\mu$Hz,
showing it in the sequence of panels on the right-hand side of Fig.\ 2.  
Following the same strategy as before, we detect five signals in this
frequency domain. After prewhitening these, we arrive at the residual
amplitude spectrum in the lowest panel of Fig.\ 2. The noise level at
frequencies below $\sim 800 \mu$Hz is artificially decreased due to our
low-frequency filtering.

Consequently, we have separately analysed our longest ($\Delta T > 2$~hr)  
light curves from the best nights, leaving them unfiltered for low
frequencies, and searched them for periodicities. We found no such
variations, with detection thresholds of 1.8 mmag for frequencies below
200~$\mu$Hz, 0.8 mmag for frequencies between 200 and 450~$\mu$Hz, and 0.5
mmag for frequencies between 450 and 1000~$\mu$Hz, respectively.

More interesting is the excess mound of amplitude at frequencies around
1500~$\mu$Hz on top of the essentially monotonically decreasing noise that
can be discerned in the lowest panel of Fig.\ 2. This is the region where
four signals have already been detected. The excess mound of amplitude may
indicated the presence of further signals or of amplitude/frequency
variability of some of the periodicities detected previously. In any case,
a preliminary frequency solution derived in this way is listed in Table 2.

We have only accepted signals that satisfy the statistical criterion of
$S/N>4$ for independent oscillations and $S/N>3.5$ for combination
frequencies. These variations are regarded as statistically significantly
detected. We refer to Breger et al.\ (1999) for a more in-depth discussion
of this criterion, that has turned out to be extremely reliable in the
past.

\begin{figure*}
\includegraphics[angle=0,width=180mm,viewport=-10 0 510 580]{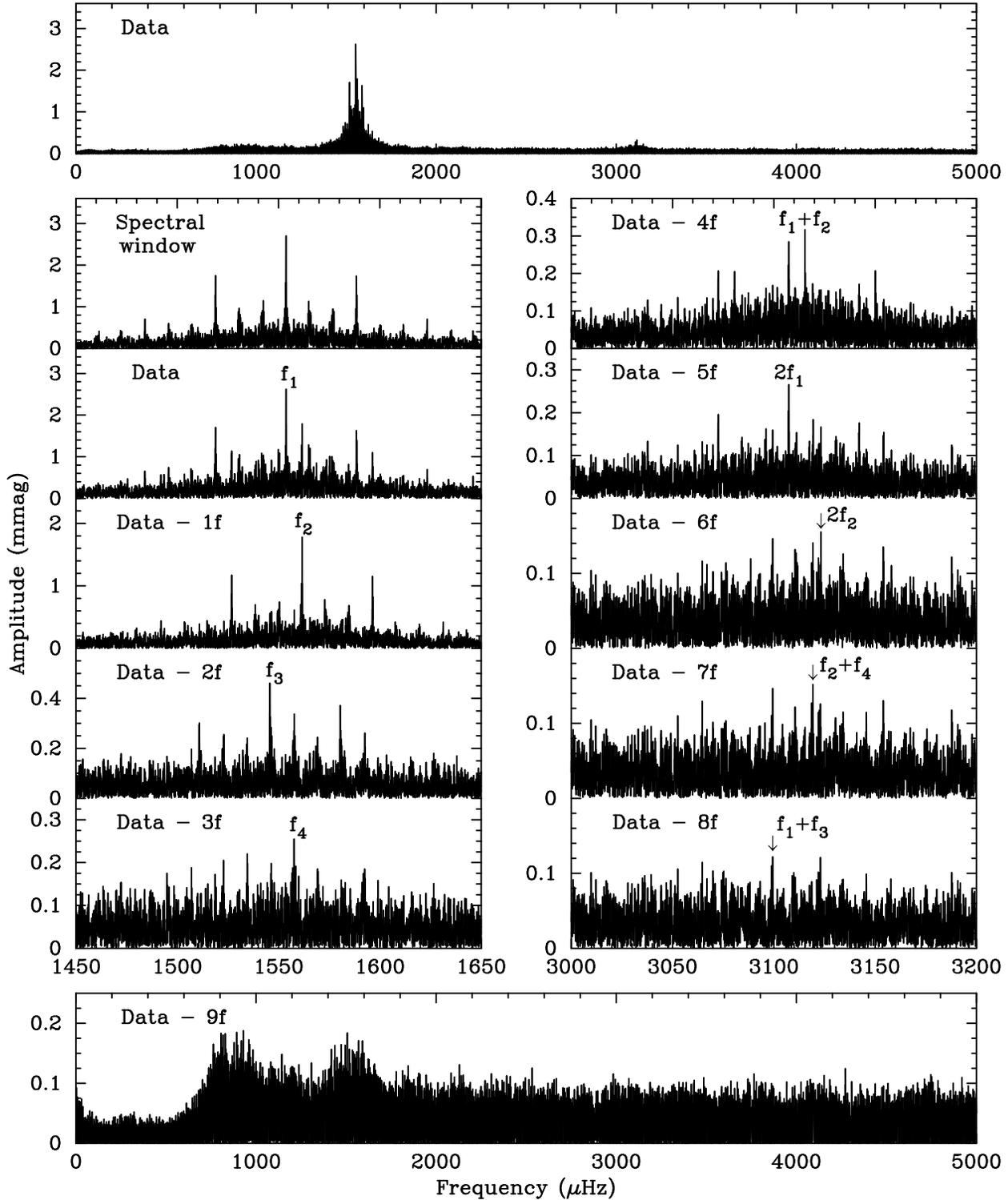}
\caption[]{Amplitude spectra of HD 99563 from our multi-site observations
with successive prewhitening of the detected frequencies.}
\end{figure*}

\begin{table}
\caption[]{Preliminary multifrequency solution for our B filter data of
the roAp star HD 99563, with the frequencies around 1500~$\mu$Hz left as
free parameters. Error estimates correspond to 1$\sigma$ values are
determined following Montgomery \& O'Donoghue (1999).}
\begin{center}
\begin{tabular}{lccc}
\hline
ID & Frequency & B Ampl. & S/N\\
 & ($\mu$Hz) & (mmag) &\\
\hline
$f_{1}$ & 1553.6780 $\pm$ 0.0007 & 2.69 $\pm$ 0.03 & 49.1\\
$f_{2}$ & 1561.6286 $\pm$ 0.0011 & 1.80 $\pm$ 0.03 & 32.9\\
$f_{3}$ & 1545.736 $\pm$ 0.004 & 0.46 $\pm$ 0.03 & 8.4\\
$f_{4}$ & 1557.639 $\pm$ 0.007 & 0.28 $\pm$ 0.03 & 5.1\\
$f_{1}+f_{2}$ or $2f_{4}$ & 3115.3065 & 0.32 $\pm$ 0.03 & 9.1\\
$2f_{1}$ & 3107.3559 & 0.25 $\pm$ 0.03 & 7.2\\
$2f_{2}$ & 3123.2571 & 0.16 $\pm$ 0.03 & 4.7\\
$f_{2}+f_{4}$ & 3119.268 & 0.14 $\pm$ 0.03 & 3.9\\
$f_{1}+f_{3}$ & 3099.414 & 0.13 $\pm$ 0.03 & 3.6\\
\hline
\end{tabular}
\end{center}
\end{table}

The independent frequencies in Table 2 can, at first sight, not be
unambiguously be interpreted within the OPM. Although $f_3, f_1$ and $f_2$
form a triplet with a spacing of about 7.95~$\mu$Hz, $f_1, f_4$ and $f_2$
can also be reconciled with a triplet with a spacing of approximately
3.98~$\mu$Hz, allowing for a $\sim 3\sigma$ shift in frequency $f_4$. In
this context it is important to note that the formal error estimates we
use here are believed to underestimate the real errors by about a factor
of 2 (Handler et al.\ 2000, Jerzykiewicz et al.\ 2005), and therefore we
cannot simply reject the latter interpretation. Consequently, all four
frequencies could even be part of an equally spaced quintuplet with a
component missing near 1549.7~$\mu$Hz.

\subsection{The mean light observations}

We again used {\tt Period98} for the frequency analysis. Amplitude spectra 
of the measurements in the five UBVRI filters were computed. Although the 
nominal Nyquist frequency of our observations is 0.5 \cd ($\sim$ 
6\,$\mu$Hz), we carefully examined a frequency range up to 25\,$\mu$Hz. We 
found variability in all the five passbands, with those in U having the 
highest amplitude, as shown in Fig.\ 3.

\begin{figure}
\includegraphics[angle=0,width=88mm,viewport=-00 5 265 245]{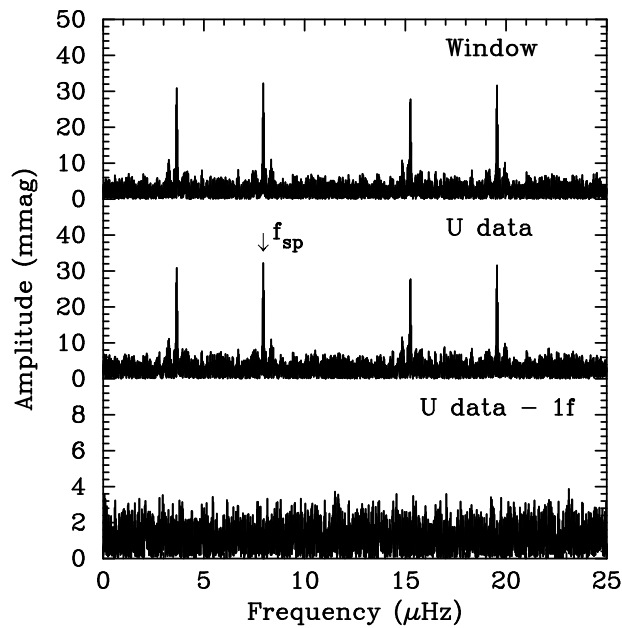}
\caption[]{Amplitude spectra of the mean light observations of HD 99563 in
the U filter. Upper panel: spectral window function of the time series.  
Middle panel: amplitude spectrum of the data themselves. Lower panel:
residual amplitude spectrum after prewhitening the strongest signal from
the data. Note the five times larger ordinate scale of the lower panel.}
\end{figure}

The upper panel of this graph contains the spectral window function of the
data, calculated as the amplitude spectrum of a single noise-free sinusoid
with a frequency of 7.95~$\mu$Hz and an amplitude of 31~mmag. There is
strong 1 \cd aliasing (the peak near 20~$\mu$Hz), and also reflection of
the aliasing pattern at frequency zero (peaks near 4 and 15~$\mu$Hz).
Nevertheless, the input signal produces the tallest peak in the window
function. This is why we can be sure that the peak labelled in the
amplitude spectrum of the data (middle panel of Fig.\ 3) is also the
correct frequency present in the measurements, which is confirmed by the
prewhitened amplitude spectrum in the lower panel of this figure that
contains noise only.

We encountered the same situation in the B filter data, where the mean 
light variations also have considerable amplitude. In $VRI$ the situation 
is more complex, as the $S/N$ of the variability is lower. Significant 
peaks at the same frequencies are however still present, but due to the 
effects of noise the tallest maxima are found at alias frequencies in $V$ 
and $R$. After prewhitening the signal near 7.95~$\mu$Hz, the residual 
amplitude spectra in all filters contain noise only; in particular, no 
evidence for harmonics or subharmonics of this frequency can be detected.

Consequently, we can describe the variability present in the mean light
variations of HD~99563 with a single frequency of 7.9498 $\pm$
0.0002~$\mu$Hz, as determined from a nonlinear least-squares fit to the U
data, which have the best signal-to-noise. This frequency is consistent
with one of the possible multiplet spacings within the pulsational signals
as determined in the previous section, but the ambiguity of whether this
frequency or half of it corresponds to the rotation period of HD~99563 is
not yet resolved. The amplitudes and phases of the mean light variations
with respect to the frequency of 7.9498~$\mu$Hz are listed in Table 3.

\begin{table}
\caption[]{The amplitudes and phases of the mean light variations of
HD~99563, once more with error estimates from Montgomery \& O'Donoghue
(1999) The phases are with respect to pulsation amplitude maximum at HJD
2452031.29627.}
\begin{center}
\begin{tabular}{ccc}
\hline
Filter & Amplitude & Phase\\
 & (mmag) & (rad)\\
\hline
U & 31.3 $\pm$ 1.0 & $-0.06 \pm 0.03$ \\
B & 20.4 $\pm$ 0.9 & $-0.00 \pm 0.04$ \\
V & 4.6 $\pm$ 0.8 & $+3.09 \pm 0.17$ \\
R & 9.2 $\pm$ 0.8 & $+3.09 \pm 0.09$ \\
I & 7.3 $\pm$ 0.7 & $+3.11 \pm 0.10$ \\
\hline
\end{tabular}
\end{center}
\end{table}

This table shows that the mean light maxima in the U and B filters occur
at the same time as pulsation amplitude maximum in the light curves. In 
the V, R and I filters, mean light minimum coincides with pulsation 
amplitude maximum. We find no statistically significant phase lag of the 
mean light extrema relative to pulsation amplitude maximum.

\section{The rotation period of HD 99563}

To infer the pulsational and magnetic geometry of HD 99563, its rotation
period has to be known with certainty. However, at this point it is not
clear whether the 7.95~$\mu$Hz signal in the mean light variations and the
frequency splitting within the pulsational signals correspond to the
rotation period or to half the rotation period.

Since the magnetic field of the star also varies with aspect, we can
invoke published magnetic field measurements for HD 99563 to resolve this
ambiguity. Hubrig et al.\ (2004) measured a longitudinal magnetic field
strength of $-688\pm145$~G on HJD\,2452494.479, whereas Kudryavtsev \&
Romanyuk (2005) measured $+580\pm100$~G on HJD\,2453395.550.

Assuming a rotation frequency of 7.9498 $\pm$ 0.0002~$\mu$Hz, the two
measurements would have been taken 618.91 $\pm$ 0.02 rotational periods
apart. Since these measurements would then have been taken at nearly the
same rotation phase, but the magnetic field showed a reversal between
these two measurements, such a rotation frequency can be ruled out. Using
half this value (3.9749 $\pm$ 0.0001~$\mu$Hz) implies that 309.455 $\pm$
0.009 rotations have gone by, which is perfectly consistent with the
magnetic field reversal between the two magnetic field observations and
the absolute magnetic field strengths measured.

The correct value of the rotation period of HD 99563 is therefore close to 
2.91\,d. We show the corresponding phase diagram of the mean light 
variations of the star in Fig.\ 4.

\begin{figure}
\includegraphics[angle=0,width=88mm,viewport=0 5 300 400]{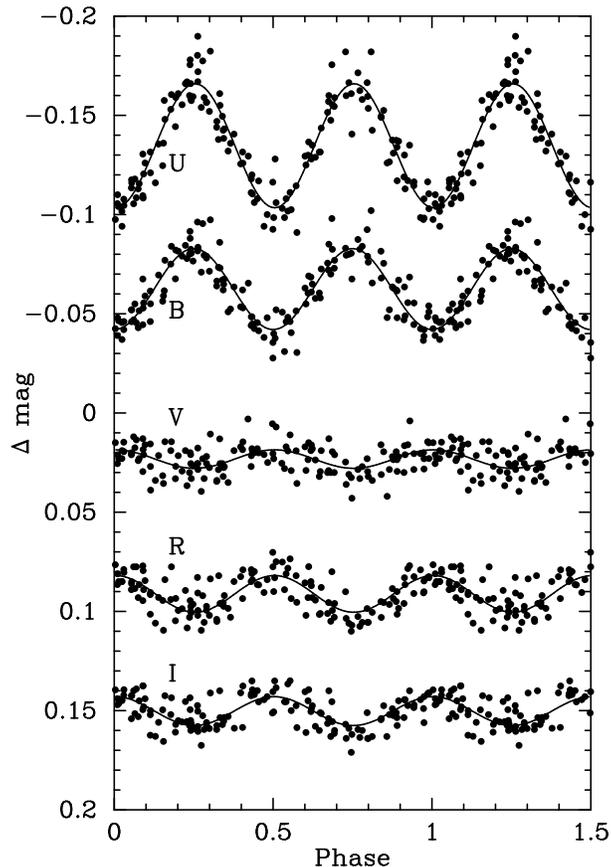}
\caption[]{Mean UBVRI light variations of HD 99563 phased with a period of 
2.91179\,d. One and a half rotation cycles are shown. The solid line is 
the single-frequency fit from Table 3.}
\end{figure}

\section{Frequency analysis (II)}

Knowing that the rotation frequency of HD 99563 is near 3.975~$\mu$Hz, we
can now determine a final multifrequency solution for the pulsation data
under the assumption of the OPM. We first examine the variations in the
frequency domain near 1550~$\mu$Hz. The spacings within all the
frequencies already detected in this range are consistent with this
rotation frequency and thus belong to the same multiplet. We therefore
searched for possible further multiplet components in the residual
amplitude spectra that have $S/N>3.5$. We indeed detected a peak near
$f_5=1569.6~\mu$Hz that is consistent with this splitting and has
$S/N>3.6$. Adding this signal to the four previously detected, we find a
symmetrical series of peaks centred on $f_4=1557.6$~$\mu$Hz, for which we
show a schematic graphical representation in Fig.\ 5.

\begin{figure}
\includegraphics[angle=0,width=88mm,viewport=0 5 280 180]{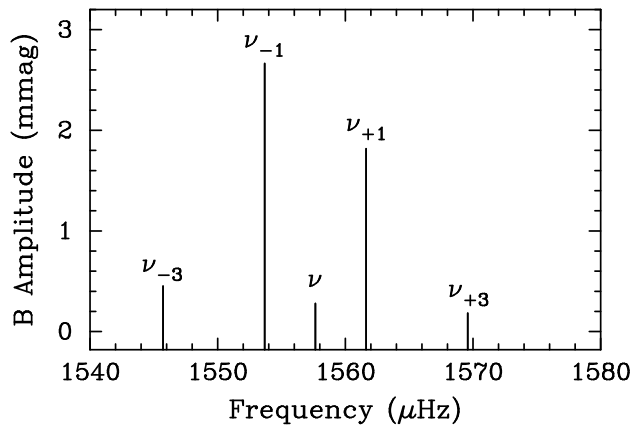}
\caption[]{The frequency quintuplet detected in the light curves of HD 
99563.}
\end{figure}

We can now derive a final frequency solution. To this end, we have started
with the frequencies from the preliminary solution in Table 2, added $f_5$
and fitted these ten signals to the data. During this process, we fixed
the signal frequencies to equal splitting, using a starting value of
3.9749~$\mu$Hz, and we also demanded that harmonics occurred at exactly
twice the ``parent'' frequency within {\tt Period98}. With this procedure,
we do not only determine final frequency values, but we also reduce the
number of free parameters in the fit to a minimum. This should guarantee 
the most reliable and most stable frequency solution, and the rotation 
frequency will also be determined most accurately that way. The final 
result of our frequency analysis can be found in Table 4.

\begin{table}
\caption[]{Final multifrequency solution for our B filter data of the roAp
star HD 99563, again with error estimates from Montgomery \& O'Donoghue
(1999; errors on frequencies and amplitudes are given in Table 2).
Pulsation phases are with respect to pulsation amplitude maximum at HJD
2452031.29627.}
\begin{center}
\begin{tabular}{lccc}
\hline
ID & Frequency & B Ampl. & Phase\\
 & ($\mu$Hz) & (mmag) & (rad)\\
\hline
$\nu$ = $f_{4}$ & 1557.6530 & 0.29 & $+2.63 \pm 0.09$\\
$\nu_{-1}$ = $f_{4}-f_{rot}$ & 1553.6779 & 2.67 & $+2.87 \pm 0.01$\\
$\nu_{+1}$ = $f_{4}+f_{rot}$ & 1561.6281 & 1.82 & $+2.83 \pm 0.01$\\
$\nu_{-3}$ = $f_{4}-3f_{rot}$ & 1545.7276 & 0.45 & $+2.82 \pm 0.06$\\
$\nu_{+3}$ = $f_{4}+3f_{rot}$ & 1569.5784 & 0.18 & $+3.00 \pm 0.15$\\
$2\nu$ & 3115.3060 & 0.32 & $-1.03 \pm 0.08$\\
$2\nu_{-1}$ & 3107.3557 & 0.25 & $-0.27 \pm 0.10$\\
$2\nu_{+1}$ & 3123.2562 & 0.17 & $-0.76 \pm 0.16$\\
$2\nu+f_{rot}$ & 3119.2811 & 0.15 & $-0.51 \pm 0.18$\\
$2\nu-4f_{rot}$ & 3099.4055 & 0.13 & $+0.55 \pm 0.21$\\
\hline
$f_{rot}$ & 3.9751 & \\
\hline
\end{tabular}
\end{center}
\end{table}

The splitting between the rotationally split components of the pulsation
mode of HD~99563 is 3.9751 $\pm$ 0.0003, which is the same as half the
frequency of the mean light variations within the errors. Since we know
that this splitting corresponds to the rotation frequency, we can fix the
latter by averaging the two results into 3.9749 $\pm$ 0.0001~$\mu$Hz,
corresponding to a rotation period of 2.91179 $\pm$ 0.00007 d.

The referee has noted that relative phases of the first-order combination
frequencies ($\phi(2\nu) - 2\phi(\nu) = 0.01 \pm 0.20$; $\phi(2\nu_{-1}) -
2\phi(\nu_{-1}) = -0.27 \pm 0.10$; $\phi(2\nu_{+1}) - 2\phi(\nu_{+1}) =
0.14 \pm 0.16$) are consistent with zero. This is an indication that the 
frequencies around 3115~$\mu$Hz are indeed the first harmonics of the main 
pulsational signals near 1557~$\mu$Hz.

\section{Discussion}

The pulsational light variations of HD 99563 are due to a single pulsation
mode at a frequency of 1557.6530~$\mu$Hz. Due to the rotation of the star
with a period of 2.91179 d, we see it at different aspect during the
rotation cycle, which gives rise to an equally spaced frequency triplet.
Two more multiplet components separated by three times the rotation
frequency were also detected. These are an indication that the pulsation
mode of HD 99563 is not a pure dipole, but is distorted by the effect of
its magnetic field.

To examine the nature of this pulsation mode, we first need to know the
inclination of the stellar rotation axis to the line of sight, which in
turn requires an estimate of the stellar radius. Consequently, we first
infer the star's position in the HR Diagram.

HD 99563 is a visual binary with a secondary component 1.2 magnitudes
fainter at a distance of 1.79$\arcsec$ (Fabricius \& Makarov 2000). Thus,
photoelectric photometric observations of the system will include the
secondary, which contributes 1/4 of the total light. (This also means that
the amplitudes of the photometric variations we discussed here must be
increased by one third to give intrinsic values.) The Tycho-2 photometry
of the two components (Fabricius \& Makarov 2000) transformed to the
standard Johnson system by using the relations by Bessell (2000) gives:  
$V=8.72, (B-V)=0.20$ for HD 99563A and $V=9.91, (B-V)=0.285$ for HD
99563B. These values suggest that the two components could be physically
associated.

The HIPPARCOS parallax of the system (ESA 1997) is $4.27\pm2.02$~mas. The
galactic reddening law by Chen et al.\ (1998) then results in
$A_v=0.010\pm0.002$. Therefore one obtains $M_v=1.9^{+0.8}_{-1.4}$ for HD
99563A and $M_v=3.1^{+0.8}_{-1.4}$ for HD 99563B. Because of the large
uncertainty of the HIPPARCOS parallax we can therefore not obtain a radius
estimate of HD 99563A with sufficient accuracy, but we can, using also
the $(B-V)$ estimated above, point out that HD 99563B is located inside
the $\delta$~Scuti instability strip and may therefore also show
pulsations. However, our check for low-frequency variability did not 
detect these.

To estimate the effective temperature of HD 99563 we can apply the
calibration by Moon \& Dworetsky (1985), who use the Str\"omgren
$H_{\beta}$ index as a temperature indicator. None of the other
Str\"omgren indices is suitable for basic parameter determination of
chemically peculiar A stars because of heavy line blanketing. The measured
$H_{\beta}$ value for HD 99563AB is 2.830 (Olsen \& Perry 1984). With the
V magnitude and $(B-V)$ colour differences given above, combined with the
standard relation by Crawford (1979), we can determine $H_{\beta}=2.844$
for HD 99563A and $H_{\beta}=2.785$ for HD 99563B. The calibration by Moon
\& Dworetsky (1985) then gives $T_{\rm eff}\sim8050$~K for HD 99563A and
$T_{\rm eff}\sim7400$~K for HD 99563B.

Elkin et al.\ (2005) performed a spectral analysis of HD~99563, which
resulted in $T_{\rm eff}=7700$~K, log $g=4.2$ and $[M/H]=0.5$ for the roAp
star. We assume a generous $T_{\rm eff}=7900\pm300$~K and that HD 99563A
is still on the main sequence, i.e. log $g=4.15\pm0.2$. Consequently, we
estimate that $R=1.9^{+0.6}_{-0.4}$~R$_{\odot}$. The projected rotational
velocity of HD 99563A is $28.5\pm1.1$~\kms (Elkin et al.\ 2005). Combined
with the rotation period determined before, we then derive
$i=60^{+30}_{-19}$ degrees, a poor constraint. However, this also implies 
$R \geq 1.58$~R$_{\odot}$.

As a final attempt to derive the basic parameters of HD 99563A, we assumed
that HD 99563B is a physical companion. Due to the wide separation of the
two components, they would have evolved independently from each other. The
goal now is to find two stellar models that have effective temperatures as
determined above (for reasons of consistency, we take the photometric
values for both stars), that have the same age, and the observed magnitude
difference of HD 99563A and B.

We used the Warsaw-New Jersey stellar evolution code (e.g.\ see Pamyatnykh
et al. 1998 for a description) to find such a pair of models. We computed
stellar evolutionary tracks in the range between 1.5 to 2.2~$M_{\odot}$
for $Z=0.02$. We indeed found two models that satisfy the observational
constraints. They have masses of 2.03~$M_{\odot}$ (HD 99563A) and
1.585~$M_{\odot}$ (HD 99563B), luminosities of 21.6 and 7.1~$L_{\odot}$
respectively, and ages of 620 Myr. The model for HD 99563A has a radius of
2.38~$R_{\odot}$, which then implies $i=43.6\pm2.1\degr$. The internal 
accuracy of this method is high: changing the secondary mass by only 
$\pm 0.01 M_{\odot}$ already cannot reproduce the observations within the 
errors.

For the purpose of further discussion, we assume $i=44\degr$. We can now
apply the OPM, which predicts
\begin{equation}
\frac{A_{+1}+A_{-1}}{A_{0}}=\tan i \tan \beta
\end{equation}
(Shibahashi 1986), where the $A_{r}$'s are the amplitudes of the $r^{th}$
rotational sidelobes, $i$ is the inclination of the stellar rotational
axis to the line of sight and $\beta$ is the magnetic obliquity. From the
amplitudes in Table 4, we obtain tan\,$i$\,tan$\beta=15.5\pm1.5$, hence
$\beta=86.4\pm0.3\degr$. This is a perfectly reasonable result given the
relative amplitudes in the central frequency triplet.

Within the framework of a regular perturbation treatment, Shibahashi \&
Takata (1993) derived an expected amplitude ratio
\begin{equation}
\frac{A_{3}+A_{-3}}{A_{2}+A_{-2}}=\frac{1}{6}\tan i \tan \beta,
\end{equation}
which is consistent with our observational data that lead to a null
results when searching for the $\nu_{2}$ and $\nu_{-2}$ components of the
stellar pulsation mode.

We further examine the distorted dipole mode of HD 99563 by applying the
axisymmetric spherical harmonic decomposition method by Kurtz (1992) which
is based on the theory by Shibahashi \& Takata (1993), to our data. This
technique breaks the magnetically distorted mode up into its pure $\ell=0,
1, 2, ...$ spherical harmonic components and consequently allows one to
infer the shape of the mode. We applied the method to our frequency
solution (Table 4), assuming zero amplitude for the unobserved $\nu_{-2},
\nu_{+2}$ components of the mode. The result is given in Tables~5 and 6.

\begin{table}
\caption[]{Components of the spherical harmonic series description of the 
pulsation mode of HD 99563 for $i=44\degr,\beta=86.4\degr$.}
\begin{center}
\begin{tabular}{lcccc}
\hline
$\ell$ & 0 & 1 & 2 & 3\\
\hline
$A_{-3}^{(\ell)}$ [mmag] & & & &    0.450\\
$A_{-2}^{(\ell)}$ [mmag] & & &    0.158 &    0.158\\
$A_{-1}^{(\ell)}$ [mmag] & &    3.405 &    0.037 &   $-$0.699\\
$A_{0}^{(\ell)}$ [mmag] &   0.237 &    0.376 &    $-$0.093 &    0.042\\
$A_{+1}^{(\ell)}$ [mmag] & &    2.372 &    0.028 &   $-$0.524\\
$A_{+2}^{(\ell)}$ [mmag] & & &    0.088 &    0.088\\
$A_{+3}^{(\ell)}$ [mmag] & & & &    0.180\\
$\phi^{(\ell)}$ [rad] &    0.024 &    2.859 &   $-$0.322 &    2.820\\
$C_{n\ell}\Omega/K^{mag}$ & &    0.107 &    0.020 &    0.010\\
\hline
\end{tabular}
\end{center}
\end{table}

\begin{table}
\caption[]{Comparison of the observed amplitudes and phases of the 
distorted dipole pulsation mode of HD 99563 and our spherical 
harmonic fit.}
\begin{center}
\begin{tabular}{lcccc}
\hline
ID & $A_{obs}$ & $A_{calc}$ & $\phi_{obs}$ & $\phi_{calc}$\\
\hline
$\nu$  & 0.29 $\pm 0.03$ & 0.29 & $+2.63 \pm 0.09$ & +2.63 \\
$\nu_{-1}$  & 2.67 $\pm 0.03$ & 2.67 & $+2.87 \pm 0.01$ & +2.87\\
$\nu_{+1}$  & 1.82 $\pm 0.03$ & 1.82 & $+2.83 \pm 0.01$ & +2.87\\
$\nu_{-2}$  & 0.00 (assumed) & 0.00 & $-----$ & n/a \\
$\nu_{+2}$  & 0.00 (assumed) & 0.00 & $-----$ & n/a \\
$\nu_{-3}$  & 0.45 $\pm 0.03$ & 0.45 & $+2.82 \pm 0.06$ & +2.82\\
$\nu_{+3}$  & 0.18 $\pm 0.03$ & 0.18 & $+3.00 \pm 0.15$ & +2.82\\
\hline
\end{tabular}
\end{center}
\end{table}

From the values in Table 5 it can be seen that the mode of HD 99563 is 
dominated by the dipole component, which is reasonable as the magnetic 
fields in Ap stars are predominantly dipoles. The contribution of the 
$\ell=0$ and $\ell=2$ terms to the mode are fairly small, but the $\ell=3$ 
component does have some influence. Table 6 shows that the observations 
are quite well reproduced by our fit. It would be interesting to see 
whether the theory by Saio \& Gautschy (2004) is capable of reproducing 
our results.

We can now check how well our model reproduces the observed pulsational
amplitudes and phases over the stellar rotation cycle. To this end, we
subdivided the time series into pieces some $4-5$ pulsation cycles long
and determined the amplitudes and phases for the 1557.653~$\mu$Hz
variation within these subsets. The spherical harmonic decomposition 
method by Kurtz (1992) yields a fit to the amplitude/phase behaviour over 
the rotation cycle, and we show it together with the data in Fig.\ 6.

\begin{figure}
\includegraphics[angle=0,width=88mm,viewport=0 5 286 310]{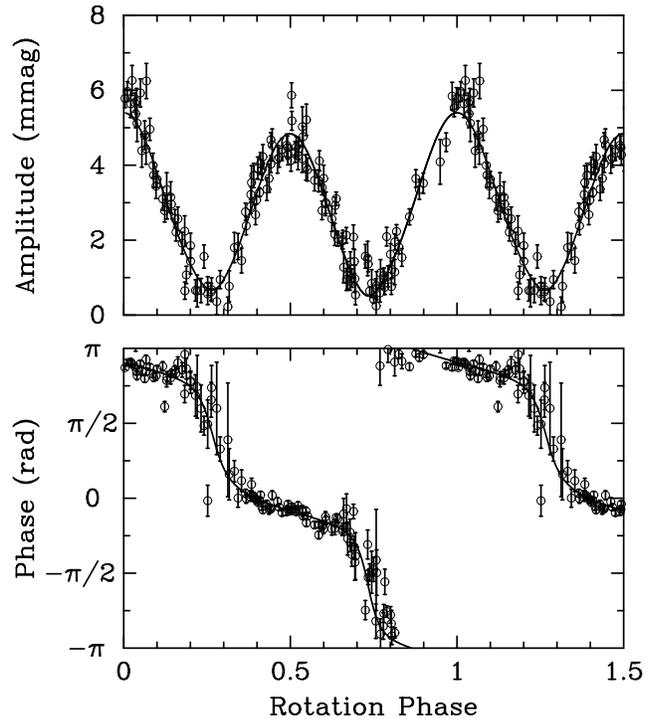}
\caption[]{Pulsational amplitudes and phases relative to a rotation period 
of 2.91179\,d. Phase zero corresponds to pulsation amplitude maximum, 
and one and a half rotation cycles are shown. The line is a fit computed 
with Kurtz' (1992) spherical harmonic decomposition method.}
\end{figure}

We can see that both magnetic (pulsation) poles are actually in view for 
approximately the same amount of time; the relative fractions are 53 and 
47 per cent. This is also the reason why it was so difficult to determine 
the rotation period of the star; the pulsations as well as the mean light 
variations are almost symmetrical along the rotation period.

The fitted curves reproduce the observations fairly well, with one
exception: the pulsation amplitude is overestimated for the pole that is
for a shorter time in the line of sight, whereas the amplitude is
underestimated for the other pole. The harmonic frequencies cannot be made
responsible for this, because they do not affect the first-order
amplitudes analysed here. Consequently, we believe that the poor fits near
the pulsation amplitude maxima could be due to presently undetected
additional multiplet components of the pulsation mode; the amount of over-
and under-fitting can be fully explained with the noise level in our
residual amplitude spectrum.

\section{Conclusions}

Our photometric multisite observations of the roAp star HD 99563 resulted
in the detection of a frequency quintuplet that is due to a single
distorted dipole pulsation mode. The splitting within this frequency
quintuplet together with our mean light observations and published
magnetic measurements allowed us to determine the stellar rotation period
as $2.91179 \pm 0.00007$~d. Within the errors, the mean light extrema
occur in phase with pulsation amplitude maximum, suggesting that the
abundance spots on HD 99563 should be fairly concentric around its
magnetic poles.

To our knowledge, HD 99563 is only the fourth roAp star where both
magnetic poles become visible throughout a rotation cycle (the others are
HR 3831, e.g., see Kurtz et al.\ (1993), HD 6532, e.g., Kurtz et al.\
(1996), and HD 80316 e.g., Kurtz et al.\ (1997)), and only the third whose
pulsational mode distortion has been quantified. In this context it is
interesting to note that the observational features of HD 99563 are in
many aspects similar to another roAp star, the well-studied HR 3831 (e.g.,
Kurtz et al.\ 1993, Kochukhov 2004): their geometrical orientations,
rotation period and effective temperatures are alike. Only the pulsation
period of HD 99563 is some 10 per cent shorter, and the phases its
first-order combination frequencies are more consistent with them being
harmonics. It would thus be very interesting to compare these two objects
in detail.

This would however require more in-depth studies of HD 99563. Its basic
parameters are still poorly known; the radius we inferred is based on the
assumption that the visual companion is physical, which needs to be
checked. If HD 99563 A and B were a physical pair, then the separation of
the components is too wide for motion around a common centre of mass to be
determined in a reasonable period of time. However, common proper motion
may be detectable within a few years.

The most efficient way to pin down the radius of the roAp star would
perhaps come from a combination of magnetic and polarisation measurements 
(Landolfi et al.\ 1997). With that, the inclination $i$ and the magnetic 
obliquity $\beta$ can be constrained, and given the accurate rotation 
period and $v \sin i$ already available, a fairly accurate radius could be 
obtained.

There are several other possibilities that make HD 99563 interesting for
future observations. As already argued by Elkin et al.\ (2005), the star
is a very attractive target for pulsational radial velocity measurements.  
Given the high amplitude of the mean light variations of the star and its
favourable geometry, Doppler Imaging of its surface should also be within
reach. Finally, we have evidence that we have not yet deciphered the full
pulsational content of the star's light curves. Another multi-site
campaign, aiming at obtaining $>200$~h of observations on 1-metre-class
telescopes would therefore also be justifiable. As we pointed out, HD
99563B is located within the $\delta$ Scuti instability strip and should
therefore also be tested for pulsations, with more suitable means than
ours.

\section*{ACKNOWLEDGEMENTS}

The Austrian Fonds zur F\"orderung der wissenschaftlichen Forschung
partially supported this work under grants S7303-AST and S7304-AST. We are
grateful to Don Kurtz for supplying results prior to publication, for
helpful discussions and for comments on a draft version of this paper.  
We also thank Hiromoto Shibahashi for his accurate referee's report.

\end{document}